\documentclass[%
 reprint,
superscriptaddress,
 amsmath,amssymb,
 aps,
]{revtex4-2}

\usepackage{CJK}
\usepackage{graphicx}
\usepackage{dcolumn}
\usepackage{bm}
\usepackage{hyperref}
\usepackage{physics}
\usepackage{macros}
\usepackage{pgfplots}

\newcommand{\rmd}{{\rm d}}

\begin{document}

\CJKnospace
\begin{CJK*}{UTF8}{gbsn}

\preprint{APS/123-QED}

\title{A New Probe of Cosmic Birefringence Using Galaxy Polarization and Shapes}

\author{\CJKfamily{bsmi} Weichen Winston Yin (尹維晨)}
\email{winstonyin@berkeley.edu}
\affiliation{Department of Physics, 366 Physics North MC 7300, University of California, Berkeley, CA 94720, USA}

\author{Liang Dai (戴亮)}
\affiliation{Department of Physics, 366 Physics North MC 7300, University of California, Berkeley, CA 94720, USA}

\author{Junwu Huang (黄俊午)}
\affiliation{Perimeter Institute for Theoretical Physics, 31 Caroline St. N., Waterloo, Ontario N2L 2Y5, Canada}

\author{Lingyuan Ji (吉聆远)}
\affiliation{Department of Physics, 366 Physics North MC 7300, University of California, Berkeley, CA 94720, USA}

\author{Simone Ferraro}
\affiliation{Physics Division, Lawrence Berkeley National Laboratory, Berkeley, CA 94720, USA}
\affiliation{Department of Physics, 366 Physics North MC 7300, University of California, Berkeley, CA 94720, USA}

\date{\today}

\begin{abstract}

We propose a novel statistical method to measure cosmic birefringence and demonstrate its power in probing parity violation due to axions. Exploiting an empirical correlation between the integrated radio polarization direction of a spiral galaxy and its apparent shape, we devise an unbiased minimum-variance estimator for the rotation angle, which should achieve an uncertainty of $5^\circ$--$ 15^\circ$ per galaxy. Large galaxy samples from the forthcoming SKA continuum surveys, together with optical shape catalogs, promise a comparable or even lower noise power spectrum for the rotation angle than in the CMB Stage-IV (CMB-S4) experiment, with different systematics.

\end{abstract}


\maketitle

\end{CJK*}


\noindent

{\it Introduction} ---
The study of fundamental discrete symmetries have led to important discoveries~\cite{Lee:1956qn,Wu:1957my,Kobayashi:1973fv}. The axion was originally proposed to solve the strong CP problem in Quantum Chromodynamics~\cite{axion1,axion2,axion3,Peccei:1977ur}. Generic axions were later shown to arise abundantly in string theory~\cite{Svrcek:2006yi, Arvanitaki:2009fg}. The Chern-Simons interaction between the axion and electromagnetism causes the parity-violating cosmic birefringence effect~\cite{Carroll:1989vb}:
\begin{equation} \label{eq:lagr}
    \mathcal{L} = - \frac{1}{4}\,g_{a\gamma\gamma}\,a \,F^{\mu\nu}\,\tilde{F}_{\mu\nu} = g_{a\gamma\gamma}\,a\,{\bf E}\cdot {\bf B},
\end{equation}
where $a$ is the axion field, $F^{\mu\nu}$ is the field strength tensor, and $\tilde{F}^{\mu\nu}$ its dual. As a photon travels, its polarization plane rotates by an angle proportional to the net change in the unwrapped axion field value along its worldline, multiplied by the axion-photon coupling $g_{a\gamma\gamma}$~\cite{Harari:1992ea,Fedderke:2019ajk,Agrawal:2019lkr}. Recent analyses of the cosmic microwave background (CMB) have shown hints of birefringence uniform across the sky at a fraction of a degree~\cite{MinamiKomatsu2020, Eskilt:2022cff}. Possible foreground contamination is being investigated~\cite{Clark:2021kze}. On the other hand, searches for anisotropic birefringence have not uncovered a signal~\cite{Gluscevic2012:WMAP7constraint,Lee:2014rpa,Contreras:2017sgi,Yin:2021kmx}.

Measuring birefringence toward any single astrophysical source is hindered by our lack of knowledge about the {\it intrinsic} polarization direction. The use of the CMB as a probe statistically solves this problem. Rotation angles correlated across the sky are measurable following accurate cosmological prediction for the Gaussian statistics of the primary CMB anisotropies~\cite{Kamionkowski2009, Yadav2009, Gluscevic2009, Yin:2021kmx}. To robustly confirm cosmic birefringence, it is important to consider alternative detection methods based on independent cosmological or astrophysical probes.

In this paper, we propose a new probe of cosmic birefringence that relies on surveying spiral galaxies with both radio polarization and apparent shape measurements. This is motivated by empirical findings that the {\it integrated} radio emission of nearby spiral galaxies at $4.8\,\mathrm{GHz}$ is significantly polarized and aligns on average with the apparent minor axis of the optical galactic disk~\cite{Stil:2008ew}. A plausible explanation for this alignment is that the polarized radio emission is predominantly synchrotron radiation powered by star formation feedback and has a polarization direction locally perpendicular to the ordered toroidal magnetic field. Integrating over an inclined disk results in a non-zero net polarization. It seems reasonable to assume that such correlation is true for a cosmological population of spirals. The shape ellipse can be measured with either optical or interferometric radio imaging, with the former available from weak lensing catalogs~\cite{Gatti2021DESshape, Giblin2021KiDS1000shape, Li2022HSCY3shape} or from imaging surveys~\cite{DESI:2018ymu, Chang:2013xja}.

\begin{figure}
    \centering
    \includegraphics[width=0.48\textwidth]{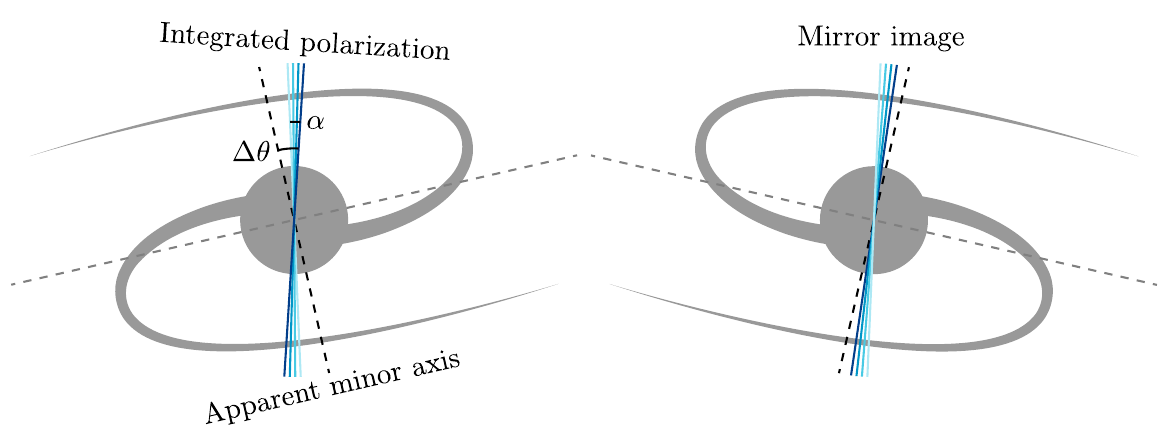}
    \caption{(Mis)alignment between the polarization of the integrated radio emission (solid blue) and the apparent minor axis (dashed black) of a spiral galaxy.}
    \label{fig:illustration}
\end{figure}

The polarization-shape alignment is imperfect for each single galaxy. Nonetheless, one expects that the mean misalignment angle is zero among unrelated galaxies, as long as all relevant physics on the (sub-)galactic scale respects parity. If there is one galaxy whose integrated polarization is offset from the apparent minor axis to the clockwise side, it should be equally probable that another galaxy appears as its mirror reflection and has a counter-clockwise misalignment (\reffig{illustration}).

It has been proposed that polarization-shape correlation can be exploited to calibrate galaxy intrinsic shapes and reduce noise in weak lensing shear measurements~\cite{BrownBattye2011, BrownBattye2011b, Whittaker2015SeparateShearIA,Francfort_2022}. Reversing this logic, one can calibrate the intrinsic polarization direction to source morphology~\cite{Kamionkowski:2010ss, Whittaker:2017hnz}. With cosmic birefringence, unrelated galaxies in a patch of the sky will have polarization rotated by nearly the same angle, leading to a mean misalignment biased from zero.

In this Letter, we construct a minimum-variance quadratic estimator for polarization rotation using galaxies with both integrated radio polarization and shape measurements, and forecast its angular noise power spectrum. Radio polarization measurements may come from the VLA Sky Survey~\cite{Lacy:2019rfe} or the upcoming SKA continuum surveys~\cite{Jarvis2015aska.confE..18J}, while the optical counterparts can come from photometric galaxy catalogs with shape measurements, such as from the DESI Legacy Surveys (DECaLS)~\cite{DESI:2018ymu} or LSST~\cite{Chang:2013xja}.

{\it Galaxy Radio Polarization} --- The polarization degree of the integrated radio emission depends on the disk inclination and radio luminosity, as well as the randomly oriented component of the magnetic field throughout the galaxy. Ref.~\cite{Stil:2008ew} found that the net polarization degree of nearby spirals at $4.8\,\mathrm{GHz}$ is $\sim 1\%$--$18\%$ with a mean $\sim 4\%$. The sample of \cite{Stil:2008ew} shows a mean misalignment angle consistent with zero and an RMS value of $\sim 26^\circ$. As we will show, the effective scatter in this misalignment angle, which limits the precision of birefringence measurements, can be reduced to $5^\circ$--$15^\circ$, once a proper estimator is used.

{\it Estimating Cosmic Birefrigence} --- Linear polarization is described by the Stokes parameters $(Q,\,U)$. Define a spin-2 variable:
\begin{equation}
    P_\pm \equiv Q\pm i\,U = I\,p\,e^{\pm 2\,i\,\theta_P},
\end{equation}
where $I$ is the intensity~\footnote{In radio astronomy, $I$ is the spectral flux density typically measured in units of jansky (Jy).}, $p$ is the polarization degree ($0 < p < 1$), and $\theta_P$ is the polarization position angle. The apparent elliptical shape of an inclined galaxy can be described by the spin-2 ellipticity~\cite[p.~11]{Kilbinger:2014cea}
\begin{equation}
    \epsilon_\pm \equiv q\pm i\,u = (a-b)/(a+b)\,e^{\pm 2\,i\,\theta_\epsilon},
\end{equation}
where $a$ and $b$ are the semi-minor and semi-major axes, respectively, and $\theta_\epsilon$ is the position angle of the semi-major axis. Since $\theta_P-\theta_\epsilon$ is rotationally invariant, we define dimensionless spin-0 quantities
\begin{equation}
    X \pm i\,Y \equiv -P_\pm\,\epsilon_\mp/I = p\,|\epsilon|\,e^{\pm 2\,i\,\Delta\theta}, \label{eq:XY_def}
\end{equation}
where $\Delta\theta \equiv \theta_P-\theta_\epsilon-\pi/2$ is the misalignment angle and $|\epsilon| \equiv (a-b)/(a+b)$. The quantities $X$ and $Y$ have even and odd parity, respectively.

The quantities $p$, $|\epsilon|$, and $\Delta\theta$ are intrinsic to each galaxy. Since there is no known evidence for {\it macroscopic} parity violation in the interstellar medium (ISM) correlated between many galaxies over intergalactic distances, we assume that $\Delta\theta$ has a distribution symmetric about zero, and that correlated parity-violating effects only occur externally as a result of cosmic birefringence, which can be quantified by a polarization rotation field $\alpha(\hatv n)$ coherent over some scales on the sky. This field may be uniform on the sky~\cite{Carroll:1998zi}, or behave as a Gaussian random field~\cite{Pospelov:2008gg, Li:2008tma}, or is highly non-Gaussian~\cite{Agrawal:2019lkr}. Thus, cosmic birefringence causes the distribution of $\Delta\theta$ to be biased from zero by an amount $\alpha(\hatv n)$ for galaxies in a local sky patch in the direction $\hatv n$. For a first analysis, we assume that $\alpha(\hatv n)$ is independent of galaxy distance, and that $(I\,p, \Delta\theta)$ are independently drawn from the same global distribution for individual galaxies.

The Faraday effect in the Milky Way (MW)'s ISM also rotates the polarization coherently on the sky, with an angle that scales with the wavelength squared. This can be distinguished from the wavelength-independent effect of the axion~\cite{Harari:1992ea}. We assume that MW Faraday rotation is corrected prior to the following analysis. This does require observation in multiple radio bands, the effectiveness of which is subject to further study. On the other hand, inhomogeneous Faraday rotation within the source galaxy causes depolarization of the integrated emission particularly at lower radio frequencies \cite{Stil:2008ew, Taylor2024MIGHTEEfields}. This is not a concern here, as depolarization at $4.8\,\mathrm{GHz}$ is included in the polarization-shape (mis)alignment data.

Parity-odd combinations of $X$ and $Y$ have a vanishing mean over the galaxy ensemble when $\alpha(\hatv n)=0$:
\begin{equation}
    \ev{Y} = \ev{X\,Y} = 0.
\end{equation}
Statistically significant deviation of $\ev{Y}$ or $\ev{XY}$ from zero would therefore be a smoking gun of parity violation.

With a non-zero $\alpha(\hatv n)$, the polarization is rotated while the disk shape is unaffected:
\begin{equation}
    P_\pm \to P_\pm\,e^{\pm 2\,i\,\alpha(\hatv n)},\quad \epsilon_\pm \to \epsilon_\pm.
\end{equation}
This implies for $X$ and $Y$
\begin{align}
    X &\to X\,\cos 2\alpha(\hatv n) - Y\,\sin 2\alpha(\hatv n),\label{eq:rotation1}\\
    Y &\to Y\,\cos 2\alpha(\hatv n) + X\,\sin 2\alpha(\hatv n).
    \label{eq:rotation2}
\end{align}
Notably, the value of $\alpha(\hatv n)$ does not change observables of a galaxy along a different direction $\hatv n'\neq \hatv n$.

We construct an unbiased, minimum-variance quadratic estimator for $\alpha(\hatv n)$ using $X$ and $Y$ following weak-lensing ~\cite{Hu_2002} and birefringence~\cite{Kamionkowski2009, Gluscevic2009, Yadav2009, Namikawa:2016fcz} applications. It can be derived via maximizing the likelihood~\cite{Hirata:2002jy} (\refapp{max_likelihood}).

For a single galaxy, it is
\begin{equation}
    \hat\alpha = \frac{\tilde\sigma_Y^2\,\overline X\,\tilde Y + (\sigma_X^2 - \sigma_Y^2)\,\tilde X\,\tilde Y}{2\left[(\sigma_X^2 - \sigma_Y^2)^2 + \tilde\sigma_X^2\,\overline X^2\right]},
    \label{eq:alpha_minvar}
\end{equation}
whose variance under the null hypothesis is
\begin{equation}
    \ev{\hat\alpha^2} = \frac{\tilde\sigma_X^2\,\tilde\sigma_Y^2}{4\left[(\sigma_X^2 - \sigma_Y^2)^2 + \tilde\sigma_X^2\,\overline X^2\right]}.
    \label{eq:var}
\end{equation}
The errors in $\tilde X$ and $\tilde Y$ including measurement noises are:
\begin{equation}
    \tilde\sigma_{X,Y}^2 = \sigma_{X,Y}^2 + f\,\left(\sigma_\epsilon/\mathrm{SNR}\right)^2, \label{eq:sigma_tilde}
\end{equation}
where $\sigma_\epsilon \equiv \sqrt{\ev{|\epsilon|^2}} \approx 0.3$ is the intrinsic galaxy shape error, and $\sigma^2_{X,Y}$ are the intrinsic variances of $X$ and $Y$. Radio interferometry with pure noise has $f=1$ (detailed derivation in \refapp{qe_derivation}). In the following, we derive numerical forecast assuming the conservative $f=1$.

{\it Sensitivity} --- From the observed $p$, $|\epsilon|$, and $\Delta\theta$, $X$ and $Y$ for each galaxy can be computed. We obtain values for $p$ and $\Delta\theta$ for 13 nearby spirals from Ref.~\cite{Stil:2008ew}, and compute values for $|\epsilon|$ using SIMBAD shapes~\cite{Wenger_2000}. We exclude M31 as its proximity and high degree of polarization ($18\%$) hint at systematics. This results in the following:
\begin{align}
    \overline{X}\pm \sigma_X &= (1.3 \pm 2.1)\times 10^{-2},\nonumber\\
    \overline{Y} \pm \sigma_Y &= (0.04 \pm 4.4)\times 10^{-3}. \label{eq:XY_stats}
\end{align}
Despite the small sample size, we find $\sqrt{\langle|\epsilon|^2\rangle}=0.30$, consistent with what is assumed in galaxy weak lensing. It is more uncertain whether the radio properties of this sample are representative of a cosmological sample. Verification would require analyzing distant galaxies (e.g. \cite{Lacy:2019rfe, Knowles2022MeerKat, Taylor2024MIGHTEEfields}).
Using values in \refeq{XY_stats} for $\overline X$, $\sigma_X$ and $\sigma_Y$ and setting $\overline Y = 0$, \refeq{alpha_minvar} gives $\langle \hat\alpha \rangle = 0.5^\circ$ and a sample standard deviation $4.4^\circ$, consistent with no birefringence; \refeq{var} forecasts $\sqrt{\ev{\hat\alpha^2}} = 5^\circ$ per galaxy in the limit of negligible measurement error.

Remarkably, $\sqrt{\ev{\hat\alpha^2}} = 5^\circ$ is five times smaller than the standard deviation of $\Delta\theta$ at $26^\circ$ as estimated from nearby spirals. We attribute this improvement to an anti-correlation between the scatter in the misalignment angle $\Delta\theta$ and the product $p\,|\epsilon|$. According to \refeq{XY_def}, when computing the statistics of $X$ and $Y$ in \refeq{XY_stats}, galaxies with large $\Delta\theta$ are down-weighted due to smaller $p\,|\epsilon|$ values.

This anti-correlation has a geometrical origin \cite{Stil:2008ew}. The intrinsic misalignment angle has a large variance for galaxies showing a low $p$ and a low $|\epsilon|$, as these low-inclination galaxies appear rather circular. Since $p$ and $|\epsilon|$ both increase with inclination, $p\,|\epsilon|$ down-weights face-on galaxies and emphasizes edge-on ones on which birefringence would be more detectable. While we do not pursue it in this work, it seems possible to accurately model this anti-correlation as a function of the inclination and derive an improved definition of $X$ and $Y$ to replace \refeq{XY_def}.

The $5^\circ$ error may result from a particularly good polarization-shape alignment in the small galaxy sample. Larger samples are required to refine this figure. Our primary claim is not this precise error value, but rather that our novel estimator effectively exploits the inclination-dependent statistics of the misalignment angle and has the potential to dramatically improve sensitivity to birefringence.

\begin{figure}
    \centering
    \includegraphics[width=0.48\textwidth]{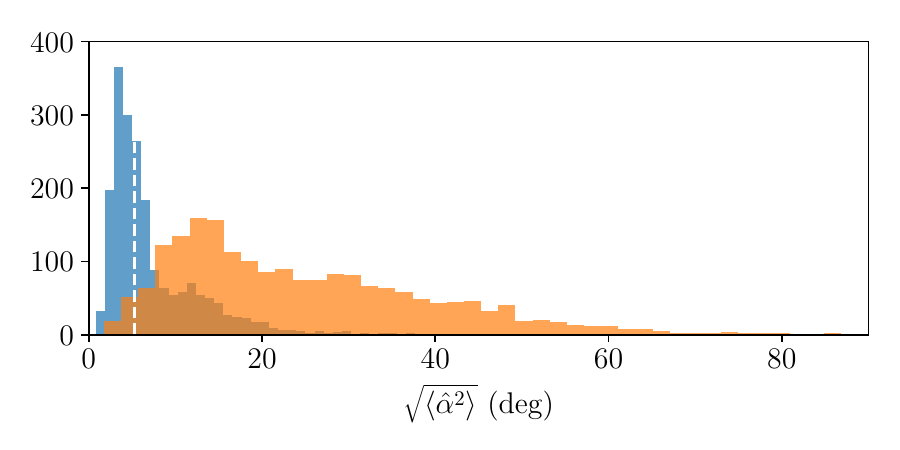}
    \caption{Distribution of the standard deviation of $\hat\alpha$ from 2000 bootstrap sets of 13 galaxies (blue), compared to the case of erasing the anti-correlation between $p|\epsilon|$ and $\Delta\theta$ (orange). The dashed line marks the standard deviation of $\hat\alpha$ measured from the Stil et al.~sample of 13 galaxies. Its lower value compared to the orange distribution indicates a genuine anti-correlation between $\Delta\theta$ and $p\,|\epsilon|$.}
    \label{fig:bootstrap}
\end{figure}

\reffig{bootstrap} estimates the uncertainty in this number by bootstrapping and demonstrates that the anti-correlation between $p|\epsilon|$ and $\Delta\theta$ is not a statistical fluke. In each run, $p|\epsilon|$ and $\Delta\theta$ are independently drawn with replacement from the Stil et al.~sample to form 13 artificial galaxies, from which $\ev{\hat\alpha^2}$ is computed. In 2,000 runs, 97.5\% produce a higher estimator uncertainty than $5^\circ$, strongly supporting a genuine anti-correlation between $\Delta\theta$ and $p\,|\epsilon|$. We note that $\sqrt{\langle\hat\alpha^2\rangle}=13^\circ$ if NGC 4565, an edge-on galaxy with $p=10.5\%$, is excluded. Its ellipticity implies a $\sim 10\%$ chance of occurrence for such ``good'' galaxies. Therefore, it may be expected that there is one such galaxy out of a sample of 13.

{\it Galaxy Survey} --- Under the flat-sky approximation, the estimator is given by a sum over $N$ galaxies
\begin{equation}
    \hat\alpha(\mathbf L) = (\Omega/N)\,\sum_{i=1}^N\,e^{-i\mathbf L\cdot\hatv n_i}\,\hat\alpha(\hatv n_i),
\end{equation}
where $\Omega$ is survey area in steradians. The estimator has a variance
\begin{equation}
    \ev{\hat\alpha(\mathbf L)\,\hat\alpha(\mathbf L')^*} = (\Omega/N)^2\,\ev{\hat\alpha^2}\sum_{i=1}^N e^{-i\,(\mathbf L-\mathbf L')\cdot\hatv n_i}.
\end{equation}
When galaxies are dense on the sky, the sum approximates a Dirac delta function:
\begin{equation}
    \ev{\hat\alpha(\mathbf L)\,\hat\alpha(\mathbf L')^*} = (\Omega/N)\,\ev{\hat\alpha^2}(2\pi)^2\,\delta^{(2)}(\mathbf L-\mathbf L').
\end{equation}
Without birefringence, the rotation estimator has a white-noise angular power spectrum
\begin{equation}
    C_{\hat\alpha\hat\alpha}(\mathbf L) = (\Omega/N)\,\ev{\hat\alpha^2}. \label{eq:white_noise}
\end{equation}

We adapt the procedure used in Ref.~\cite{Bonaldi_2016} to count available spiral galaxies. We first compute the redshift-dependent radio luminosity function of spiral galaxies. The radio luminosity function at $4.8\,\mathrm{GHz}$ at redshift $z$ can be related to that at $1.4\,\mathrm{GHz}$ at $z=0$, taking into account luminosity evolution \cite{Negrello_2007} and the spectral index ($S_\nu \propto \nu^\alpha$ where $\alpha=-0.7$ \cite{Sadler_2002, Condon_2002}). We choose $4.8\,\mathrm{GHz}$ because Ref.~\cite{Stil:2008ew} established the empirical correlation at this frequency and Faraday rotation in both the source and the MW is greatly reduced at higher frequencies. Next, the local radio luminosity function can be obtained from the far-infrared (FIR) luminosity function \cite{Takeuchi_2003,Takeuchi_2004} using a nearly linear relation between the radio and FIR luminosities of star-forming galaxies~\cite{Helou:1985mr,Kewley_2002,Condon1992,Yun_2001,Garrett2002,De_Zotti_2009}.

The number of spirals with an observed spectral flux density $S_\nu$ above a threshold $S_\nu^{\mathrm{min}}$ is
\begin{align}
    & N[S_\nu>S_\nu^{\mathrm{min}}] =\label{eq:N_galaxies}\\
    &\quad\Omega\int \rmd z \int_{\log \lambda(z)^{-1}L_{4.8}^{\mathrm{min}}(z)}^{\log L_{4.8}^{\mathrm{max}}}\,\rmd \log L_{4.8}\, \frac{c\,s(z)^2}{H(z)}\,\phi(\log L_{4.8}),\nonumber
\end{align}
where $s(z) = \int_0^z c/H(z')\,\rmd z'$ is the comoving distance to redshift $z$; $H(z)$ is the expansion rate for $H_0=70\,\mathrm{km\,s^{-1}\,Mpc^{-1}}, \Omega_\Lambda = 0.7, \Omega_m = 0.3$; $\Omega$ is the surveyed sky area in steradians; $\phi$ is the $4.8\,\mathrm{GHz}$ radio luminosity function (per unit comoving volume) of spirals at $z=0$ (see \refapp{luminosity_function}); $L_{4.8}^{\mathrm{max}} = 10^{25}\,\mathrm{W\,Hz^{-1}}$ is a generous upper bound on the radio luminosity of a spiral galaxy; $L_{4.8}^{\mathrm{min}}$ is the galaxy luminosity at redshift $z$ with an observed spectral flux density $S_\nu^{\mathrm{min}}$:
\begin{equation}
    L_{4.8}^{\mathrm{min}}(z) = 4\pi\,s(z)^2\,(1+z)^{1-\alpha}\,S_\nu^{\mathrm{min}};
\end{equation}
for the radio spectral index $\alpha=-0.7$~\cite{Sadler_2002, Condon_2002}, and $\lambda(z)$ describes the luminosity evolution with $z$ for spiral galaxies~\cite{Negrello_2007}. The upper limit in the $z$ integral should not exceed $z=1.5$; above that, nearly all massive halos host spheroidal galaxies~\cite{Granato_2004,Silva:2004zza}. The lower limit in the $z$ integral should exceed the redshift of detectable birefringence sources. There is a trade-off between the galaxy number and the accumulation of birefringence along the line of sight. A more sophisticated calculation considering birefringence sources interspersed amongst the galaxies throughout all redshifts is model-dependent and is left for a future study.

The highest sensitivity achievable with data in the near future will likely come from correlating polarized SKA radio galaxies with LSST shape data (or using resolved SKA radio shapes). Similar to Ref.~\cite{coogan2023looking}, we define \texttt{Wide}, \texttt{Deep}, and \texttt{Ultradeep} reference surveys, which have $\sigma_I = 1\,\mathrm{\mu Jy}$, $0.2\,\mathrm{\mu Jy}$, and $0.05\,\mathrm{\mu Jy}$, and cover $1000\,\mathrm{deg}^2$, $20\,\mathrm{deg}^2$, and $1\,\mathrm{deg}^2$, respectively, and consider SKA-MID Band 5a ($4.6$--$8.5\,$GHz~\cite{Braun2019arXiv191212699B}). Including galaxies whose integrated $4.8\,\mathrm{GHz}$ intensities have ${\rm SNR}>5$, we estimate the number of measurable spiral galaxies at $0.2<z<1.5$ to be $N=1.6\times 10^6$, $2.1\times 10^5$ and $2.8\times 10^4$ for each survey, respectively. For $0.8<z<1.5$, the numbers are $N=2.8\times 10^5$, $9.5\times 10^4$ and $1.7\times 10^4$, respectively (see \refapp{luminosity_function} for calculation details).

For most detectable galaxies the polarization measurement error significantly contributes to the variance of the estimator $\hat\alpha$. Those galaxies collectively provide a non-negligible amount of information to reduce estimation noise. Consider $N$ galaxies in a small sky patch $\rmd\Omega$, each with a rotation estimator $\hat\alpha_i$. A linear combination $\hat\alpha_{\mathrm{best}} \equiv \sum^N_{i=1} w_i\,\hat\alpha_i$ with weights $w_i = \ev{\hat\alpha_i^2}^{-1}/\sum_j\ev{\hat\alpha_j^2}^{-1}$ gives an unbiased minimum-variance estimator. For different galaxies, $\ev{\hat\alpha_i^2}$ only depends on the unpolarized radio SNR, which in turn only depends on $z$ and $L_{4.8}$. Let us denote it as $\ev{\hat\alpha^2}(x)$, where $x=(z,\,\log L_{4.8})$.

With \refeq{N_galaxies} rewritten as $N = \rmd\Omega\,\int \Phi(x)\,\rmd x$, the weights are
\begin{equation}
    w(x) = \rmd\Omega^{-1}\, \ev{\hat\alpha^2}(x)^{-1}\,\left[ \int \ev{\hat\alpha^2}(x')^{-1} \Phi(x')\,\rmd x'\right]^{-1},
\end{equation}
where $\Phi(x)$ quantifies the galaxy distribution with redshift and luminosity as defined in \refeq{appD_Phi} in \refapp{var_eff}. This leads to
\begin{equation}
    \ev{\hat\alpha_{\mathrm{best}}^2} = \rmd \Omega^{-1}\left[\int\ev{\hat\alpha^2}(x)^{-1}\Phi(x)\,\rmd x\right]^{-1}.
\end{equation}
Accounting for the SNR distribution is therefore equivalent to having $N$ identically measured galaxies, each with an effective estimator variance
\begin{equation}
    \ev{\hat\alpha^2}_{\mathrm{eff}} = N\ev{\hat\alpha_{\mathrm{best}}^2} = \frac{\int\Phi(x)\,\rmd x}{\int\ev{\hat\alpha(x')^2}^{-1}\Phi(x')\,\rmd x'}.\label{eq:var_eff}
\end{equation}
We replace $\ev{\hat\alpha^2}$ in \refeq{white_noise} with $\ev{\hat\alpha^2}_{\rm eff}$ to estimate the effective noise power spectrum
\begin{equation}
    C_{\hat\alpha\hat\alpha}(\mathbf L) = \left[\int\ev{\hat\alpha(x)^2}^{-1}\Phi(x)\,\rmd x\right]^{-1}.
\end{equation}
While the $x$-integration formally depends on the choice of the threshold $S_\nu^{\mathrm{min}}$ as in \refeq{N_galaxies}, the integral approaches a minimal value as $S_\nu^{\mathrm{min}}$ decreases, until saturation at $S_\nu^{\mathrm{min}}/\sigma_I \lesssim 5$.

\begin{figure}
    \centering
    \includegraphics[width=0.47\textwidth]{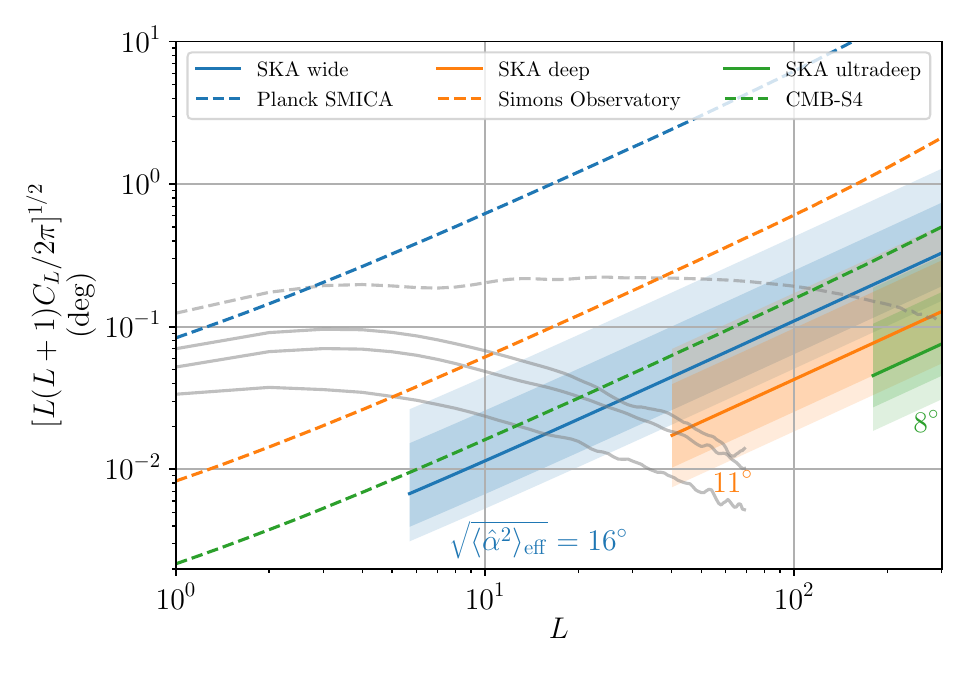}
    \caption{Polarization rotation noise power spectra for SKA radio galaxy surveys combined with an overlapping optical imaging galaxy survey, compared to those for three CMB surveys. The effective per-galaxy sensitivity (\refeq{var_eff}) is labeled in matching color. The low-$L$ cutoffs of SKA curves are set by the proposed survey areas.  We show signal power spectra from ultralight axion strings (see main text) up to $z=z_{\mathrm{cmb}}$ (gray dashed) and up to $z=1.5,\,1,\,0.5$ (gray solid; from top to bottom). Colored bands mark the $68\%$ and $95\%$ confidence intervals of the noise floor estimated from bootstrapping the sample of 13 galaxies from \cite{Stil:2008ew}.}
    \label{fig:sensitivity}
\end{figure}

\reffig{sensitivity} shows the noise angular power spectrum of the rotation estimator, compared with sensitivity curves for current and projected CMB surveys~\cite{Yin:2021kmx}. SKA will enable a noise power spectrum comparable to or even lower than that of CMB-S4 for $L\gtrsim 5$, but probe a more limited $z$ range of birefringence sources. The noise spectra are compared to the maximal signal allowed by Planck~\cite{Yin:2021kmx} due to ultralight axion strings~\cite{Agrawal:2019lkr} approximated by the loop crossing model~\cite{Jain_2021}.

{\it Discussion} --- Using \refeq{alpha_minvar}, rotation can be estimated at discrete galaxy locations $\{\hat\alpha_i\}_i$, which is suited for real-space approaches that exploit the non-Gaussian statistics of $\alpha(\hatv n)$. This is applicable to birefringence from axion strings \cite{Yin:2023vit, hagimoto2023measures}.

The reduced intrinsic noise ($5^\circ$--$15^\circ$) of our carefully constructed estimator compared to the spread of the misalignment angle ($26^\circ$) is promising for reducing shot noise and bias due to intrinsic alignment in cosmic shear estimation~\cite{BrownBattye2011}.

Measuring birefringence in the CMB has the advantage of being sensitive to all sources up to $z\approx 1100$. Spiral galaxies do not effectively probe birefringence sources at $z\gtrsim 1.5$. Nonetheless, galaxies present an opportunity for extracting the $z$-dependence of the birefringence sources. One could compare CMB rotation estimation with that from radio galaxies, or compare galaxy estimation between different $z$ bins. For example, if axion strings exist for much of the cosmic history, this would enable the measurement of their size and number evolution with $z$ \cite{Jain_2021}. An axion field that is homogeneous but dynamic for $z\lesssim 1100$~\cite{Carroll:1998zi} could also show up as $z$-dependent from galaxy-based tomographic estimation, complementing recently proposed tomography methods for the CMB~\cite{Sherwin:2021vgb,Nakatsuka:2022epj,Lee:2022udm,Namikawa:2023zux}. Furthermore, any detection of birefringence through the CMB~\cite{MinamiKomatsu2020} can be validated with galaxies. Foreground systematics , \emph{e.g.}~from the MW, would show up in every redshift bin and could thus be identified and subtracted.

{\it Conclusion} --- We have developed an unbiased minimum-variance quadratic estimator to detect anisotropic cosmic birefringence from correlation between radio polarization and shape in spiral galaxies. This offers an independent technique that complements CMB probes. This will be enabled by a synergy between upcoming large radio galaxy surveys such as SKA and optical imaging surveys such as LSST. The noise floor of the polarization rotation angular power spectrum of the galaxy-based method using SKA will be better than that of CMB-S4. We plan to test the empirical correlations of \cite{Stil:2008ew} on cosmological galaxy samples using data from existing smaller radio surveys \cite{Lacy:2019rfe, Knowles2022MeerKat, Taylor2024MIGHTEEfields}.

\acknowledgments
We thank Neal Dalal, Kangning Diao, Marc Kamionkowski, Dustin Lang, Christopher McKee, Jessie Muir, Kendrick Smith, Eiichiro Komatsu, Adrian Lee, Uro\v s Seljak, and Zack Li for useful discussions and comments.
L.D. acknowledges research grant support from the Alfred P. Sloan Foundation (Award Number FG-2021-16495), from the Frank and Karen Dabby STEM Fund in the Society of Hellman Fellows, and from the Office of Science, Office of High Energy Physics of the U.S. Department of Energy (Award Number DE-SC-0025293).
S.F. is supported by Lawrence Berkeley National Laboratory and the Director, Office of Science, Office of High Energy Physics of the U.S. Department of Energy under Contract No.\ DE-AC02-05CH11231. Research at Perimeter Institute is supported in part by the Government of Canada through the Department of Innovation, Science and Economic Development Canada and by the Province of Ontario through the Ministry of Economic Development, Job Creation and Trade.
\bibliography{refs}

\onecolumngrid
\appendix

\section{Quadratic estimator}
\label{app:qe_derivation}

Our goal is to estimate the polarization rotation field $\hat\alpha(\mathbf n)$ using $X$ and $Y$. Since $X$ and $Y$ are computed from the ellipticity, one may be concerned about weak lensing effects. For CMB anisotropies, quadratic estimators for cosmic shear and for polarization rotation are orthogonal to each other to first order in each of the effects due to their opposite parity~\cite{Yin:2021kmx}. We expect this first-order non-contamination to hold for our birefringence estimator. Cosmic shear induces a white noise on $Y \approx p\,|\epsilon|\,\gamma = 0.002\,(p/0.1)\,(\gamma/0.02)\,|\epsilon|$ where $\sqrt{\langle\gamma^2\rangle} \approx 0.02$ at $z\sim 1$~\cite{Krause:2009yr}, which is smaller than the intrinsic $\sigma_Y=0.004$ (\refeq{XY_stats}). With both weak lensing and galaxy intrinsic shape alignment, a correlated noise contribution to $\hat\alpha$ arises, but with a magnitude significantly smaller than the noise floor derived here. We will therefore disregard weak lensing below.

We first comment on the measurement error in polarization. The important quantities for our purpose are $\sigma_I$, the measurement error in the total intensity, and $\sigma_{Q,U}$, the measurement error in the Stokes parameters $Q$ and $U$, assumed to be equal. For a monochromatic electromagnetic wave described by complex electric field amplitudes $E_x,E_y$, the Stokes parameters are given by
\begin{align}
    I &= |E_x|^2 + |E_y|^2,\label{eq:stokesI}\\
    Q &= |E_x|^2 - |E_y|^2,\label{eq:stokesQ}\\
    U &= 2\Re[E_xE_y^*],\label{eq:stokesU}\\
    V &= -2\Im[E_xE_y^*].\label{eq:stokesV}
\end{align}
In the absence of any signal, we model components of $E_x,E_y$ as independent Gaussian distributions with zero mean and equal variance. The variances of the Stokes parameters satisfy
\begin{equation}
    \sigma_Q^2 = \sigma^2_U = \sigma^2_V = f\,\sigma^2_I,
\end{equation}
with $f=1$.

This derivation applies to coherent monochromatic noise. While $E_x,E_y$ are Gaussian, the Stokes parameters are non-Gaussian. In the practical case of integration over a frequency band, \refeqs{stokesI}{stokesV} are modified to statistical averages over frequency and time. The distributions of the Stokes parameters $I,Q,U,V$ approach Gaussian by the central limit theorem. In other words, the measurement noise of the Stokes parameters in a realistic measurement of radio polarization may be assumed to be Gaussian.

In practice, the presence of a signal predominant in $I$ increases $\sigma^2_I$ more than $\sigma^2_{Q,U,V}$. The precise error in radio polarization measurements varies depending on the beam profile, frequency, duration of observation, number of pointings, and whether model fitting is involved in polarization detection, generally with $f \lesssim 1$. For measurements of spiral galaxies by VLA, this ratio is less than $1/2$ (\emph{e.g.}~0.45 at $4.8\,\mathrm{GHz}$ and 0.36 at $1.4\,\mathrm{GHz}$ \cite{Beck_2007}). In forecasting sensitivities of SKA-MID configurations, we adopt $f=1$, leading to a conservative forecast. Note that Ref.~\cite{loi2019simulations} assumes $f=1/2$ for SKA-MID at $1.4\,\mathrm{GHz}$.

Next, we assume that the uncertainty in $p=P/I=\sqrt{Q^2+U^2}/I$ is dominated by uncertainty in the numerator rather than the denominator, as the unpolarized intensity typically has a much higher SNR than the polarized emission. For realistic deep optical imaging surveys with shape measurements, the radio polarization measurement error dominates over the error in ellipticity measurement, an assumption we adopt. Polarization and shape noise components are assumed to be statistically independent from each other and from the polarization and noise signals.

With tildes denoting observed quantities with measurement noise, for each galaxy with total flux $I$,
\begin{align}
    \ev{\tilde Y^2} &= \ev{[\Im(-\tilde P_+\,\tilde\epsilon_-/\tilde I)]^2}\nonumber\\
    &\approx \ev{[\Im(-\tilde P_+\,\epsilon_-/I)]^2}\nonumber\\
    &= \ev{(\tilde U^2\,\epsilon_1^2 + \tilde Q^2\,\epsilon_2^2 - 2\,\tilde Q\,\tilde U\,\epsilon_1\,\epsilon_2)/I^2}\nonumber\\
    &= \ev{Y^2} + (\sigma_U^2\ev{\epsilon_1^2} + \sigma_Q^2\ev{\epsilon_2^2}) / I^2\nonumber\\
    &= \ev{Y^2} + f\,(\sigma_I^2/I^2)\ev{|\epsilon|^2}.
\end{align}
After a similar calculation for $\ev{\tilde X^2}$, we obtain the errors in $X$ and $Y$ including measurement error:
\begin{equation}
    \tilde\sigma_{X,Y}^2 = \sigma_{X,Y}^2 + f\,\frac{\sigma_\epsilon^2}{\mathrm{SNR}^2},
\end{equation}
where the total flux SNR enters the expression, and $\sigma_\epsilon = \sqrt{\ev{|\epsilon|^2}} \approx 0.3$ as in the main text.

We now calculate the propagation of this error into the variance of the quadratic estimator. Let $\hat\alpha$ for each galaxy be constructed from the observed quantities (with tilde) including measurement errors, up to quadratic combinations:
\begin{equation}
    \hat\alpha \equiv u\,\tilde X + v\,\tilde Y + u'\,\tilde X^2 + v'\,\tilde Y^2 + w\,\tilde X\,\tilde Y,
\end{equation}
where the coefficients are to be determined.

Under a polarization rotation $\alpha$ at the galaxy's location, the observed quantities are transformed:
\begin{equation}
    \tilde X \to \tilde X - 2\alpha\,Y,\quad \tilde Y \to \tilde Y + 2\alpha\,X,
\end{equation}
where the small angle approximation ($\alpha \ll 1$) is used, since large birefringence angles are observationally ruled out. Note that the transformation is proportional to quantities without measurement noises, since cosmic birefringence is applied before detection. By requiring that the estimator $\hat\alpha$ be unbiased, namely $\ev{\hat\alpha} = \alpha$, we have
\begin{align}
    0 &= u\ev{X} + u'\ev{X^2} + v'\ev{Y^2}\\
    \frac 12 &= v\ev{X} + w\ev{X^2} - w\ev{Y^2}.
\end{align}
Using these, $u$ can be expressed in terms of $u'$ and $v'$, and $v$ can be expressed in terms of $w$. It remains to minimize the variance of the estimator, $\ev{\hat\alpha^2}$, under the null hypothesis, with respect to $u'$, $v'$, and $w$.

To enable analytic calculations, we approximate $X$ as a Gaussian random variable with mean $\overline X$ and variance $\sigma_X^2$, and $Y$ a Gaussian random variable with {\it zero} mean and variance $\sigma_Y^2$. Since $X$ and $Y$ are only approximately Gaussian, this gives a nearly optimal quadratic estimator. The true optimal estimator could be obtained by numerically minimizing $\ev{\hat\alpha^2}$ given empirical distributions for $Ip$, $|\epsilon|$, and $\Delta\theta$.

After some algebra, we obtain the claimed result in \refeq{alpha_minvar}:
\begin{equation}
    u = u' = v' = 0,\quad v = \frac{\tilde\sigma_Y^2\,\overline X}{2\left[(\sigma_X^2 - \sigma_Y^2)^2 + \tilde\sigma_X^2\,\overline X^2\right]},\quad w = \frac{\sigma_X^2 - \sigma_Y^2}{2\left[(\sigma_X^2 - \sigma_Y^2)^2 + \tilde\sigma_X^2\,\overline X^2\right]}.
\end{equation}

\section{Maximum likelihood approach}
\label{app:max_likelihood}

Inspired by the idea originally put forth in Ref.~\cite{Hirata:2002jy} and closely following the formalism presented in Ref.~\cite{Yin:2021kmx}, we pursue an maximum-likelihood approach to estimating the polarization rotation field $\alpha(\hatv n)$. For detailed justifications of some assumptions we are about to make, we refer to Sections 3 and 4.1 of Ref.~\cite{Yin:2021kmx}.

Let $\mathbf X_i \equiv \begin{pmatrix}\tilde X_i&\tilde Y_i\end{pmatrix}^T$ denote the observed quantities including measurement error for the $i$-th galaxy, located in the direction $\hatv n_i$, among a total of $N$ galaxies. We start by assuming the following log-likelihood function:
\begin{equation}
    \ln\mathcal L[\alpha;\{\mathbf X_i,\mathbf N_i\}] = -\frac 12 \sum_{i=1}^N \left\{\left[M(\alpha(\hatv n_i))^{-1}(\mathbf X_i-\mathbf N_i) - \overline{\mathbf X}_i \right]^T \,C_{\mathbf X_i}^{-1}\, \left[M(\alpha(\hatv n_i))^{-1}(\mathbf X_i-\mathbf N_i) - \overline{\mathbf X}_i\right] + \mathbf N_i^T C_{\mathbf N_i}^{-1} \mathbf N_i\right\},
    \label{eq:logL}
\end{equation}
where $\mathbf N_i$ is the measurement error of $\mathbf X_i$; $\overline{\mathbf X}_i$ is the mean of $\mathbf X$ over an ensemble of galaxies; $C_{\mathbf X_i}$ is the covariance matrix of the intrinsic, unrotated quantities $\mathbf X_i$ (which in general may depend on the type or other properties of the $i$-th galaxy); $C_{\mathbf N_i}$ is the covariance matrix of the measurement noise given by the second term in \refeq{sigma_tilde}, meaning $C_{\mathbf N_i} = (\sigma_\epsilon^2/2)\,\mathrm{SNR}_i^2\operatorname{diag}(1,1)$; and $M(\alpha(\hatv n_i))$ is the effect of polarization rotation on the intrinsic quantities. \refeq{logL} assumes that each $\mathbf X_i$ is a random Gaussian variable, the same approximation we used in the main text, meaning $C_{\mathbf X_i} = \operatorname{diag}(\sigma_{X_i}^2,\sigma_{Y_i}^2)$. Our goal is then to obtain an estimator $\hat\alpha_{\mathrm{ML}}(\hatv n)$ which maximizes $\ln\mathcal L$ given a set of observations $\{\mathbf X_i\}$.

Since the sum in \refeq{logL} has completely independent terms, it suffices to individually maximize each term:
\begin{equation}
    \ln\mathcal L_i[\alpha(\hatv n_i);\mathbf X_i,\mathbf N_i] = -\frac 12 \left[ M(\alpha(\hatv n_i))^{-1}(\mathbf X_i - \mathbf N_i) - \overline{\mathbf X}_i \right]^T\,C_{\mathbf X_i}^{-1}\, \left[ M(\alpha(\hatv n_i))^{-1}(\mathbf X_i - \mathbf N_i) - \overline{\mathbf X}_i \right] - \frac 12 \mathbf N_i^T C_{\mathbf N_i}^{-1} \mathbf N_i.
\end{equation}
From now on, we will omit the galaxy index $i$ and suppress the $\hatv n_i$ dependence for clarity. For any fixed $\mathbf X$, the likelihood is maximized when the noise is
\begin{equation}
    \mathbf N_{\mathrm{max}} = (\mathbf M^{-T}C_{\mathbf X}^{-1}\mathbf M^{-1} + C_{\mathbf N}^{-1})^{-1} \mathbf M^{-T}C_{\mathbf X}^{-1} (\mathbf M^{-1}\mathbf X - \overline{\mathbf X}).
\end{equation}
The value of the likelihood at this noise realisation is
\begin{equation}
    \ln\mathcal L[\alpha;\mathbf X,\mathbf N=\mathbf N_{\mathrm{max}}] = -\frac 12 (\mathbf X-\mathbf M^{-1}\overline{\mathbf X})^T (C_{\mathbf N} + \mathbf M^T C_{\mathbf X}\mathbf M)^{-1} (\mathbf X - \mathbf M^{-1}\overline{\mathbf X}).\label{eq:logL_i}
\end{equation}
It would have been more justifiable to marginalize over the noise $\mathbf N$, which would produce an additional logarithmic term that is numerically negligible.

The matrix $M(\alpha)$ depends on the angle $\alpha$ non-linearly. Using \refeqs{rotation1}{rotation2}, we expand it up to $O(\alpha^2)$:
\begin{equation}
    M(\alpha) = e^{2\,\alpha\, T} = 1 + 2\,\alpha\, T + 2\,\alpha^2\, T^2,\quad T = \begin{pmatrix}
        0&1\\
        -1&0
    \end{pmatrix}.
\end{equation}
We insert this expansion into \refeq{logL_i} and take the derivative with respect to $\alpha$, keeping terms up to linear order in $\alpha$:
\begin{equation}
    \frac{\partial\mathcal L}{\partial\alpha} = \frac{2}{\tilde\sigma_X^2\,\tilde\sigma_Y^2}\left[\tilde\sigma_Y^2\,\overline X\,\tilde Y + (\sigma_X^2-\sigma_Y^2)\,\tilde X\,\tilde Y\right] - \frac{4\,\alpha}{\tilde\sigma_X^2\,\tilde\sigma_Y^2}\left[\tilde\sigma_Y^2\,\overline X\tilde X + (\sigma_X^2-\sigma_Y^2)\left(\tilde X^2 - \tilde Y^2\right)\right].
\end{equation}
Setting this to zero results in the maximum-likelihood estimator
\begin{equation}
    \hat\alpha_{\mathrm{ML}} = \frac{\tilde\sigma_Y^2\,\overline X\,\tilde Y + (\sigma_X^2-\sigma_Y^2)\,\tilde X\,\tilde Y}{2\left[\tilde\sigma_Y^2\,\overline X\tilde X + (\sigma_X^2-\sigma_Y^2)\left(\tilde X^2 - \tilde Y^2\right)\right]}.
\end{equation}
We then introduce the approximation by replacing the denominator of the estimator with its ensemble average (expectation value):
\begin{equation}
    \hat\alpha_{\mathrm{ML}} = \frac{\tilde\sigma_Y^2\,\overline X\,\tilde Y + (\sigma_X^2-\sigma_Y^2)\,\tilde X\,\tilde Y}{2\left[\tilde\sigma_X^2\,\overline X^2 + (\sigma_X^2-\sigma_Y^2)^2\right]}.
\end{equation}
The quadratic estimator in \refeq{alpha_minvar} is thus recovered. When $\alpha=0$, the variance $\ev{\hat\alpha_{\mathrm{ML}}^2}$ is therefore also given by \refeq{var}.

\section{Effective estimator variance}
\label{app:var_eff}

\refeq{var} and \refeq{sigma_tilde} tell us the dependence of the estimator error $\sqrt{\ev{\hat\alpha^2}}$ on the RMS flux $\sigma_I$ of a given radio survey. For a cosmological sample of galaxies, we can use the redshift-dependent radio luminosity function of galaxies to obtain a distribution of values for $\ev{\hat\alpha^2}$ when measurement errors are accounted for.

We wish to include as many galaxies as possible in our estimation of $\alpha$, even if many may have low SNR for the polarized flux. We now calculate the lowest estimator variance achievable by including all galaxies observable by an experiment with a given $\sigma_I$.

Suppose we are estimating the birefringence $\alpha$ within a small solid angle $\rmd\Omega$ around a direction $\hatv n$. The galaxies located within this patch of sky follow the redshift-dependent radio luminosity function. That is, in a redshift bin from $z$ to $z+\rmd z$ and a luminosity bin from $\log L$ to $\log L + \rmd\log L$, the number of galaxies in a solid angle $\rmd \Omega$ is
\begin{equation}
\label{eq:appD_Phi}
    \rmd N = \Phi(z,\log L)\,\rmd z\,\rmd\log L\,\rmd\Omega.
\end{equation}
Each of these galaxies corresponds to an independent estimator $\hat\alpha_i$ whose variance $\ev{\hat\alpha(p)^2}$ only depends on $p=(z,\log L)$ of the galaxy. We define a new estimator as a linear combination of them:
\begin{equation}
    \hat\alpha_{\mathrm{best}} \equiv \sum_i w_i\,\hat\alpha_i.
\end{equation}
The weights are required to sum to unity:
\begin{equation}
    1 = \sum_i\,w_i.
\end{equation}
We assume that the weight $w_i = w(p)$ only depends on $p$. In the continuum limit, we have
\begin{equation}
    1 = \rmd\Omega\,\int w(p)\,\Phi(p)\,\rmd p.
\end{equation}
Under this constraint, we choose weights that minimise the variance of $\hat\alpha_{\mathrm{best}}$:
\begin{equation}
    \ev{\hat\alpha_{\mathrm{best}}^2} = \sum_{ij} w_iw_j\ev{\hat\alpha_i\hat\alpha_j} = \sum_i w_i^2 \ev{\hat\alpha_i^2}.
\end{equation}
In the continuum limit, this can be written as
\begin{equation}
    \ev{\hat\alpha_{\mathrm{best}}^2} = \rmd\Omega\,\int w(p)^2 \ev{\hat\alpha(p)^2}\Phi(p)\,\rmd p.
\end{equation}
A standard application of the Lagrange multiplier method leads to
\begin{equation}
    w(p) = \rmd \Omega^{-1}\,\left[\int \ev{\hat\alpha(p')^2}^{-1}\,\Phi(p')\,\rmd p'\right]^{-1}\,\ev{\hat\alpha(p)^2}^{-1}.
\end{equation}
The minimum variance is
\begin{equation}
    \ev{\hat\alpha_{\mathrm{best}}^2} = \rmd \Omega^{-1}\,\left[\int \ev{\hat\alpha(p')^2}^{-1}\,\Phi(p')\,\rmd p'\right]^{-1}.
\end{equation}
This is the best estimator variance achieved from a total of $N$ galaxies following the luminosity function, where
\begin{equation}
    N = \rmd\Omega\,\int \Phi(p)\,\rmd p.
\end{equation}
If there were instead $N$ galaxies with the same estimator variance $\ev{\hat\alpha^2}_{\mathrm{eff}}$, the variance of the arithmetic mean of their corresponding estimators would be $\ev{\hat\alpha^2}_{\mathrm{eff}} / N$. What would $\ev{\hat\alpha^2}_{\mathrm{eff}}$ have to be so that the arithmetic mean would produce the same variance as $\hat\alpha_{\mathrm{best}}$? The answer is
\begin{equation}
    \ev{\hat\alpha^2}_{\mathrm{eff}} = N\,\ev{\hat\alpha_{\mathrm{best}}^2} = \frac{\int \Phi(p)\,\rmd p}{\int \ev{\hat\alpha(p)^2}^{-1}\Phi(p)\,\rmd p},
\end{equation}
which is correctly independent of the solid angle $\rmd \Omega$.

We can use this $\ev{\hat\alpha^2}_{\mathrm{eff}}$ to estimate the noise floor of the power spectrum:
\begin{equation}
    C_{\hat\alpha\hat\alpha} = \frac{\Omega}{N_{\mathrm{tot}}}\,\ev{\hat\alpha^2}_{\mathrm{eff}} = \frac{\Omega}{N_{\mathrm{tot}}} \frac{\int \Phi(p)\,\rmd p}{\int \ev{\hat\alpha(p)^2}^{-1}\,\Phi(p)\,\rmd p} = \left[\int \ev{\hat\alpha(p)^2}^{-1}\,\Phi(p)\,\rmd p\right]^{-1},
\end{equation}
which is independent of the sky area surveyed.

\section{Radio luminosity function}
\label{app:luminosity_function}

In this Appendix, we outline our procedure for estimating the source count of spiral galaxies for radio continuum surveys, thus providing the calculation details underlying \refeq{N_galaxies}.

The number of galaxies observable within a solid angle $\rmd\Omega$ of the sky, in a redshift interval between $z$ and $z+\rmd z$ and in a logarithmic interval of intrinsic spectral luminosity between $\log L$ and $\log L + \rmd \log L$, can be written as
\begin{equation}
    \rmd N = \frac{c\,s^2(z)}{H(z)}\,\phi(z,\log L)\,\rmd\Omega\,\rmd z\,\rmd\log L,\label{eq:dN}
\end{equation}
where the luminosity function $\phi(z,\log L)$ gives the number density (per unit comoving volume) of galaxies per decade of luminosity at redshift $z$, and $c\,s(z)^2/H(z)\,\rmd\Omega\,\rmd z$ is the comoving volume corresponding to the solid angle $\rmd \Omega$ and the redshift slice from $z$ to $z+\rmd z$. We need to know the functional form of $\phi$ for radio-emitting spiral galaxies.

The radio galaxy luminosity function $\phi$ at $z>0$ can be related to that at $z=0$:
\begin{equation}
    \phi(z,\log L) = \phi(0,\log[\lambda(z)L]),
\end{equation}
where the factor $\lambda(z)$ accounts for the redshift evolution of spiral galaxies. For this, we adopt the simple formula from Ref.~\cite{Negrello_2007} that fits observed counts:
\begin{equation}
    \lambda(z) = \begin{cases}
        (1+z)^{-1.5},&0\leq z <1\\
        2^{-1.5},&1\leq z< 1.5
    \end{cases}.
\end{equation}
The number density of spiral galaxies appear to drop quickly above $z=1.5$ \cite{Granato_2004,Silva:2004zza}, so we set $\phi(z,\log L)=0$ for $z \geq 1.5$.

Suppose the radio luminosity function $\phi_\nu(z,\log L)$ at some frequency $\nu$ is known. The function at a different frequency $\nu'$ can be derived as:
\begin{equation}
    \phi_{\nu'}(z,\log L) = \phi_\nu(z,\log[(\nu'/\nu)^{-\alpha}L]),
\end{equation}
by exploiting the power-law frequency dependence of the spectral flux density with respect to frequency: $S_\nu \propto \nu^\alpha$, where the spectral index has an appropriate value $\alpha=-0.7$ \cite{Sadler_2002, Condon_2002} (the readers should distinguish between this power-law index and the polarization rotation angle). While \cite{Stil:2008ew} reported the empirical polarization-shape correlation at $4.8\,$GHz, we would like to relate sources at $4.8\,$GHz to sources at $\nu = 1.4\,$GHz in order to make contact with observed counts.

For local star-forming galaxies, it is known that the radio and far infrared (FIR) luminosities are nearly linearly related \cite{Helou:1985mr,Kewley_2002,Condon1992,Garrett2002}. Consider a general mean relation between the FIR luminosity and the radio luminosity at $z=0$:
\begin{equation}
    L_{1.4} = f(L_{\mathrm{FIR}}).
    \label{eq:radio_FIR}
\end{equation}
The corresponding luminosity functions at $z=0$ are related by
\begin{equation}
    \phi_{1.4}(0,\log L_{1.4}) = \frac{L_{1.4}}{f^{-1}(L_{1.4})\,f'(f^{-1}(L_{1.4}))}\,\phi_{\mathrm{FIR}}(f^{-1}(\log L_{1.4})),
\end{equation}
where $f^{-1}$ is the inverse function to $f$. This reduces to a simple shift in the logarithmic luminosity when $f$ is linear.

The FIR luminosity function is given by a phenomenological Schechter model \cite{Takeuchi_2003}:
\begin{equation}
    \phi_{\mathrm{FIR}}(\log L_{\mathrm{FIR}}) = \phi_* \left(\frac{L_{\mathrm{FIR}}}{L_*}\right)^{1-a} \exp{-\frac{1}{2\,\sigma^2}\left[\log(1 + \frac{L_{\mathrm{FIR}}}{L_*})\right]^2}.
\end{equation}
The parameter values are taken from Eq.~(6) in \cite{Takeuchi_2003}, which correspond to the ``cool'' galaxies, which are interpreted as star-forming spiral galaxies. Corrections from \cite{Takeuchi_2004} are applied, assuming $h=0.7$:
\begin{equation}
    \phi_* = 0.635\,\mathrm{Mpc^{-3}},\quad a = 1.25,\quad L_* = 1.95\times 10^9\,L_\odot,\quad \sigma = 0.5.
\end{equation}

Following Eq.~(16) in \cite{De_Zotti_2009}, we find that the following fitting function for \refeq{radio_FIR} in combination with the FIR luminosity function above reproduces the appropriate curve in Fig.~8 of \cite{De_Zotti_2009}:
\begin{equation}
    L_{1.4} = A\left[\left(\frac{L_{\mathrm{FIR}}}{L_b}\right)^{-3.1} + \left(\frac{L_{\mathrm{FIR}}}{L_b}\right)^{-1}\right]^{-1},
\end{equation}
where the fitting parameters are given as
\begin{equation}
    A = 7.66\times 10^{20}\,\mathrm{W\,Hz^{-1}},\quad L_b = 3.34\times 10^{35}\,\mathrm{W}.
\end{equation}

Having found the redshift-dependent radio luminosity function $\phi_\nu(z,\log L)$ for any radio frequency $\nu$, we can now compute the total source count within a sky patch by integrating \refeq{dN}:
\begin{equation}
    N = \Omega\int\rmd z\int_{L_{\mathrm{min}}}^{L_{\mathrm{max}}}\rmd\log L\,\frac{c\,s(z)^2}{H(z)}\,\phi_{\nu(z)}(z,\log L),
\end{equation}
where $\nu(z) = \nu_0\,(1+z)$ is the intrinsic frequency at the location of a galaxy at redshift $z$ corresponding to an observed frequency $\nu_0$. If the radio continuum survey has an RMS flux of $S_\nu^{\mathrm{min}}$, then the luminosity cutoff is given by
\begin{equation}
    L_{\mathrm{min}} = 4\pi\,s(z)^2\,(1+z)\,S_\nu^{\mathrm{min}}.
\end{equation}
Writing $\phi(\log L) \equiv \phi_{\nu_0}(0,\log L)$ and performing a change of variables on $\log L$ from $1.4\,$GHz to $4.8\,$GHz, we derive \refeq{N_galaxies} in the main text.

\end{document}